\documentclass[12pt]{article}
\usepackage{amssymb,amsmath,framed}
\usepackage[dvips]{graphicx}

\catcode `\@=11
\def\@cite#1#2{#1\if@tempswa , #2\fi}
\catcode `\@=12

\newtheorem{theorem}{Theorem}%

\title{Fractal structure of a solvable lattice model} 
\author{{\sc Kazuhiko MINAMI} \\ \\
  {\small{\it Graduate School of Mathematics, Nagoya University,}}\\
  {\small{\it  Nagoya, 464-8602, JAPAN}}}

\begin{document} 
\maketitle  
Fractal structure of the six-vertex model is introduced   
with the use of the IFS (Iterated Function Systems). 
The fractal dimension satisfies 
an equation written by the free energy 
of the six-vertex model.   
It is pointed out that 
the transfer matrix method 
and the $n$-equivalence relation introduced in lattice theories 
have also been introduced in the area of fractal geometry. 
All the results can be generalized 
for the models suitable to the transfer matrix treatment, 
and hence this gives general relation 
between solvable lattice models and fractal geometry. 
\\
\vspace{2.4cm}

\vspace{0.3cm}

\noindent
E-mail address: minami@math.nagoya-u.ac.jp\\
Tel.+81-52-789-5578\\
Fax.+81-52-789-2829
\newpage


\setcounter{equation}{0}


\section{Introduction}

Lieb[\cite{L1}-\cite{L3}] and Sutherland[\cite{S}] 
solved the six-vertex model which is a lattice model 
now known to be equivalent to the XXZ quantum spin chain.   
The six-vertex model satisfies the Yang-Baxter equation 
which plays fundamental role in integrable lattice models.[\cite{J}, \cite{JM}] 

Halsey et al.[\cite{HJKPS}] analyzed the spectrum of singularities 
lying upon possibly fractal sets, 
for example those found in dynamical systems theory. 
This was generally formulated by Edgar and Mauldin[\cite{EM}] 
as a theory of new sub-structure lying in fractal sets. 
This sub-structure is called "multifractal". 

In this paper, 
the fractal structure lying in the six-vertex model 
is formulated as a fractal generated by an iterated function system (IFS). 
We find that the notions which correspond to the transfer matrix theory 
and an equivalence relation of boundary conditions in lattice systems 
have also been introduced in the area of fractal geometry. 
The functional relation for the fractal dimension of the fractal set 
corresponding to the six-vertex model 
shows structural change in the thermodynamic limit, 
depending on an anisotropy parameter. 

This correspondence can be formulated for systems 
suitable to the transfer matrix treatment.  
This gives us a possibility 
to find relations between theories of fractal sets and those of lattice models.  
In particular,
models which satisfy the Yang-Baxter relation provide fractal sets 
that can be mapped to integrable systems.  
This is especially interesting when we consider the fact that 
there exist infinite hierarchy of solvable models[\cite{AKW}, \cite{KAW}] 
though we do not have sufficiently many examples of fractal sets 
where we can obtain the fractal dimensions exactly. 
The relation is also interesting from the point of view 
that the six-vertex model is governed by the quantum group.[\cite{JM}]
The correspondence itself is also valid if one considers nonintegrable cases.

This note is arranged as follows. 
\ref{six-vertex} is a short review of the six-vertex model 
and the equivalence relation that classifies the free energy. 
\ref{IFS fractal} is also a short review of the IFS. 
Informed readers in respective fields may skip these sections. 
The correspondence between the transfer matrix method and IFS, 
the $n$-equivalence and the transitivity condition of IFS fractals, 
and the relation between the fractal dimension and the free energy 
of the six-vertex model is explained in \ref{result}.

\section{The six-vertex model and the n-equivalence relation}
\label{six-vertex}

Let us introduce the six-vertex model.  
We consider a rectangular lattice with $h$ rows and $w$ columns, 
and assign an arrow to each bond with the rule that 
two of four arrows point in and other two point out at each site (Fig.1(a)).  
Then six types of local arrow arrangements are possible. 
The site together with the four arrows around it is called the vertex.  
Each vertex is assumed 
to have finite and non-negative energies $\epsilon_1$, $\epsilon_2$ or $\epsilon_3$. 
The vertex energies are assumed to be invariant 
under the reversal of all arrows. 

When a line is drawn on each bond that points down or points left, 
a one-to-one correspondence 
between the configurations of arrows 
and the configurations of lines on the lattice can be found (Fig.1(a) and (b)). 
These lines do not intersect each other.   
Each line begins from one bond on the boundary,   
and continues until it reaches another bond on the boundary.  
Thus the number of lines on the lattice 
is determined by the number of lines on the boundary.  

The free energy of the six-vertex model on the rectangle was first obtained 
with the cyclic boundary conditions in two directions.[\cite{L1}-\cite{S}]
The transfer matrix $V$ of the six-vertex model 
and the Hamiltonian ${\cal H}_{\rm XXZ}$ of the XXZ quantum spin chain commute, 
and hence they share the same eigenvectors. 
There also exists a simple relation 
between the eigenvalues of $V$ and those of ${\cal H}_{\rm XXZ}$ 
(see for example [\cite{JM}]). 


Next let us introduce an equivalence relation of boundary conditions 
which is called the $n$-equivalence.[\cite{n-eq}] 
Let us consider models in which each of the local variables 
takes one of a finite number of discrete states. 
We consider the lattice 
where the number of boundary sites $N'$ 
is of a lower order than the total number of sites $N$ 
in the thermodynamic limit: $N'=o(N)$ $(N\to\infty)$. 

Let us consider a site on the lattice 
and the number of steps (the number of bonds) $n'$ 
that is necessary to reach one of the sites on the boundary. 
There exists the minimum of $n'$ for each site. 
Then let us consider the sites 
where the minimum of $n'$ is equal to $n$. 
We call them the $n$-boundary sites. 
Let us consider the set of bonds 
between $(n-1)$- and $n$-boundary sites, 
and call them the $n$-boundary bonds. 
The set of $n$-boundary sites together with the $n$-boundary bonds 
is called the $n$-boundary, 
and configurations on the $n$-boundary 
are called the $n$-boundary configurations.

Let us introduce a set of $n$-boundary configurations $\{ \Gamma_i\}$ 
which is the set of all the allowed configurations on the $n$-boundary 
under a specific boundary condition $\Gamma$ on the actual boundary of the lattice. 
Then the equivalence of boundary conditions is defined as follows: 
boundary conditions $\Gamma$ and $\Gamma'$ 
are $n$-equivalent if $\{ \Gamma_i\}=\{ \Gamma'_i\}$ 
as a set of $n$-boundary conditions.

In some models such as the six-vertex model, 
the free energy depend on the boundary condition 
still in the thermodynamic limit. 
However it is derived[\cite{n-eq}, \cite{n-eq_6V}] that 
the free energies with boundary conditions $\Gamma$ and $\Gamma'$ 
are identical in their thermodynamic limit,  
if the boundary conditions $\Gamma$ and $\Gamma'$ 
are $n$-equivalent with a finite $n$ throughout the limit. 
It is also true 
when the number $n$ diverges 
but satisfies $n=o(N/N')$ $(N\to\infty)$.

With the use of this equivalence, 
it is derived[\cite{n-eq_6V}] that 
the free energy of the six-vertex model on domain $D$,  
with fixed density of lines $\rho_1$ for the horizontal bonds  
and fixed density of lines $\rho_2$ for the vertical bonds on the boundary,  
are identical to each other in the thermodynamic limit: $f=f(\rho_1, \rho_2)$. 

The equivalence can be introduced on a part of the boundary, 
for example on the first row of the rectangle.  
In this paper, we will introduce this type of $n$-equivalence. 
In this case the corresponding $n$-boundary sites 
are those on the $(n+1)$-th row of the lattice. 

The $n$-equivalence corresponds to the irreducibility of the transfer matrix, 
and corresponds to the regularity of the stochastic matrix. 
It will be noted later that 
the $n$-equivalence also corresponds to the transitivity condition 
in fractal geometry.

\section{Iterated Function System (IFS)}
\label{IFS fractal}

Next let us introduce the iterated function system (IFS), 
which provides a way to construct fractal sets  
through iterations of contractions 
defined by a set of functions. 
First we will consider the Cantor set 
which is one of the simplest fractal sets generated by an IFS. 
Let us consider the interval $I_0=[0, 1]$  
and introduce two contractions defined by the functions 
$F_1(x)=\frac{1}{3}x$ and $F_2(x)=\frac{2}{3}+\frac{1}{3}x$.  
The functions $F_1$ and $F_2$ generate two similar subsets 
with the contraction ratio $r=1/3$: 
$F_1([0, 1])=[0, 1/3]$ and $F_2([0, 1])=[2/3, 1]$.   
Thus one obtains $I_1=F_1(I_0)\cup F_2(I_0)=[0, 1/3]\cup [2/3, 1]$.  
Beginning from $I_1$ and again operating $F_1$ and $F_2$, 
one obtains 
\begin{eqnarray}
I_2
&=&F_1(I_1)\cup F_2(I_1)\\
&=&[0, 1/9]\cup [2/9, 1/3]\cup [2/3, 7/9]\cup [8/9, 1]. 
\end {eqnarray}
Operating the functions $F_1$ and $F_2$ iteratively, 
one  obtains the subset $I_n$ 
\begin{equation}
I_n=F_1(I_{n-1})\cup F_2(I_{n-1}). 
\end {equation}
Taking the limit $n\to\infty$, 
it is known that there remains a non-vanishing subset, 
which we call the Cantor set. 
The Cantor set has been generated by the set of functions $\{F_1, F_2\}$. 
This procedure can be generalized 
to the case of the set of finite number of functions $\{ F_1, F_2, \ldots, F_p\}$, 
which we call the iterated function system. 

One can also assume that 
an additional index $j$ is assigned to each interval. 
Let us introduce contraction functions $F_{ij}$, 
which operates only on the intervals of type $j$ 
and generates the intervals of type $i$.  
Not all the intervals can be generated from the interval of type $j$: 
some of the generations $j\mapsto i$ may be prohibited. 
This restriction is often displayed through the graph (Fig.2).  
Each arrow from $i$ to $j$ 
is usually written as a symbol for the function $F_{ij}$. 
Two or more arrows from $i$ to $j$, or from $i$ to $i$, 
with different contraction ratios may exist. 
This kind of restricted IFS is called the graph-directed IFS.[\cite{B}-\cite{E}]

In the case of the Cantor set, 
2 similar small intervals are generated from an interval in each iteration 
with the contraction ratio $r=1/3$. 
In the case of the simple equal division of a $d$-dimensional interval, 
the number of generated small intervals should be $(1/r)^d$. 
Then the similarity dimension $d_{\rm S}$ is introduced 
through the relation 
\begin{equation}
2=(1/r)^{d_{\rm S}}, 
\end{equation}
and one obtains $d_{\rm S}=\log 2/\log 3=0.6309\cdots< 1$. 
The similarity dimension $d_{\rm S}$, 
which is introduced in the case of self-similar sets, 
is the simplest example of fractal dimensions.

One can introduce another fractal dimension. 
Let $E$ be a non-empty and bounded subset of ${\bf R}^d$. 
Let $N_\delta(E)$ be the smallest number of 
$d$-dimensional intervals ($d$-dimensional boxes) of diameter $\delta$, 
with which one can cover the set $E$. 
Then the box-counting dimension of $E$ is defined as 
\begin{eqnarray}
{\rm dim_B} E=\lim_{\delta\to 0}\frac{\log N_\delta(E)}{-\log \delta}. 
\end{eqnarray}
This definition means that 
the smallest number of boxes to cover $E$ 
is of order $(1/\delta)^{d_{\rm B}}$, where $d_{\rm B}={\rm dim_B} E$, 
in the limit $\delta\to 0$. 

The Hausdorff dimension of $E$, 
usually written as $d_{\rm H}={\rm dim_H} E$, 
is defined by introducing a countable collection of open sets  
to cover $E$, instead of the set of boxes. 
Taking the limit $\delta\to 0$, 
where $\delta$ is the suprimum of the diameters of the open sets, 
we can introduce the Hausdorff measure of $E$, 
and the Hausdorff dimension is defined as the dimension 
where the measure jumps from $\infty$ to $0$  
(see details for example in [\cite{F3}]). 
The Hausdorff dimension might be the most sophisticated dimension 
to measure fractal. 

In the case of complicated fractal sets, 
fractal dimensions often take different values from each other. 
However in our case, fractal sets generated by IFS, 
$d_{\rm H}$ is equal to $d_{\rm B}$. 
In the case of self-similar sets, 
we have $d_{\rm H}=d_{\rm B}=d_{\rm S}$.

\section{Fractal structure of the six-vertex model}
\label{result}

Let us consider the six-vertex model on a rectangle 
with $w$ columns and $h$ rows. 
When we fix the line configuration on the first row of the rectangle, 
we see that not all of the configurations are possible on the next low, 
because of the six-vertex restriction (Fig.1(b)). 
The set of allowed configurations 
are determined by the configuration on the first low. 
Let us assume that the configuration is type $j$ on the first low, 
and that the configurations $i_1$, $i_2$, \ldots, $i_{p_2}$ 
are allowed on the second low. 
Then assuming the configuration $i_l$ on the second low, 
configurations $i'_1$, $i'_2$, \ldots, $i'_{p_3}$ 
are allowed on the third low. 
The set of allowed configurations $\{i'_1, i'_2, \ldots, i'_{p_3}\}$ 
is determined by the configuration $i_l$. 
The configurations are generated by operating  
the row to row transfer matrix $V$. 
This iteration procedure corresponds 
to the generation of fractal sets 
by means of the graph-directed IFS. 
This is the simple summary of the procedure 
to relate lattice models and IFS fractals. 


Now let us strictly define the graph-directed IFS 
and a possible set of fractals corresponding to the six-vertex model. 

Let us consider a finite number of "dots" labeled by index $j$. 
Let us introduce a set of directed edges, 
where each edge $e_{ij}^{(k)}$ starts a dot $i$ and ends at a dot $j$.  
A pair of dots $i$ and $j$ may be joined by several edges 
distinguished by the index $k$. 
Edges from dot $i$ to $i$ itself may also exist. 
Let us introduce a contraction function $F_{ij}^{(k)}$: ${\bf R}^d\to{\bf R}^d$,    
corresponding to each $e_{ij}^{(k)}$. 
Let $r_{ij}^{(k)}$ be the contraction ratio of $F_{ij}^{(k)}$,  
which is the infimum of the number $r$ 
that satisfies $|F_{ij}^{(k)}(x)-F_{ij}^{(k)}(y)|\leq r|x-y|$ 
for all $x, y\in{\bf R}^n$. 
It is assumed that $0<r_{ij}^{(k)}< 1$. 

Let us consider a set of $n'$ directed edges 
$(e_{ik_1}, e_{k_1k_2}, \ldots, e_{k_{n'-1}j})$, $k_l\neq i, j$ $(l=1, \ldots, n'-1)$, 
which form a sequential path from $i$ to $j$. 
Let ${\cal E}^{(n')}_{ij}$ be the set of such sequential $n'$ edges. 
We assume the transitivity condition,
i.e. there is a positive integer $n$ 
which satisfies that, for all $i$, $j$, there exists an integer $n_{ij}\leq n$
such that ${\cal E}^{(n_{ij})}_{ij}$ is not empty. 
The transitivity condition means 
that there exist finite sequential paths in the graph
joining every pair of dots $i$ and $j$. 

Then it is known that 
there exists a unique family of non-empty fractal sets $\{E_j\}$ 
such that
\begin{eqnarray}
E_i=\bigcup_{j} \bigcup_{k} F_{ij}^{(k)}(E_j). 
\label{it}
\end{eqnarray}
The set of functions $\{F_{ij}^{(k)}\}$ is called a graph-directed IFS, 
and the fractal sets $\{E_j\}$ are called
a family of graph-directed sets.

When the right-hand side of (\ref{it}) is disjoint, 
we say that the set of functions $\{F_{ij}^{(k)}\}$ 
satisfies the (strong) separation condition.

In the case of the six-vertex model, 
each "dot" labeled $j$ corresponds to the set of line configurations 
allowed under the condition that the boundary line configuration on the first row 
is fixed and labeled $j$. 
Each directed edge corresponds to an allowed generation of a configuration,  
from a configuration $j$ on a row of vertical bonds   
to a configuration $i$ on the next row.

The functions $\{F_{ij}^{(k)}\}$ are introduced as follows. 
Let us introduce an order of bonds as shown in Fig.3  
and assign numbers $\{s_l\}$, 
where $s_l=+1$ (or $s_l=0)$ 
if the arrow on the $l$-th bond points down/left (or up/right). 
When we assume $w$ columns and $h$ rows 
and the cyclic boundary condition in the horizontal direction, 
the number of bonds is equal to $2wh+w$. 

The set of numbers $\{s_1, \ldots, s_w\}$ 
is determined by the line configuration on the first $w$ vertical bonds, 
which is the boundary condition on the first row. 
Let us introduce the number $j=\sum_{l=1}^w s_l 2^{-l}$, 
which works 
as an index for the configuration on the first row of vertical bonds.   
The correspondence from $\{s_1, \ldots, s_w\}$ to $[j, j+2^{-w}]$ 
provides a mapping from an allowed configuration to an interval. 

When we consider the next $2w$ bonds, we will find a row of $w$ vertices. 
Let us introduce contractions of each interval $[j, j+2^{-w}]$ as  
\begin{eqnarray}
[j, j+2^{-w}]\mapsto\bigcup_{x_2} [x_2, x_2+2^{-3w}{\tilde r}_{ij}],
\end{eqnarray}
where 
$x_2=j+\sum_{l=w+1}^{3w} s_l 2^{-l}$ 
provides an index of the line configuration on the lattice 
up to the second row of vertical bonds, 
${\tilde r}_{ij}$ is an weight which satisfies $0<{\tilde r}_{ij}< 1$ 
and will be fixed later in (\ref{bratio}), 
and in the union $\bigcup_{x_2}$ 
the index $x_2$ runs over all the allowed line configurations 
on the added $2w$ bonds 
with fixed $\{s_1, s_2, \ldots, s_w\}$. 
Let us introduce a function 
\begin{eqnarray}
x=\sum_{l=1} s_l 2^{-l}\mapsto p(x, h)=\sum_{l=(h-1)w+1}^{hw} s_l 2^{-(l-(h-1)w)},
\end{eqnarray}
which is an index 
that distinguishes line configuration $\{s_{(k-1)w+1}, s_{(k-1)w+2}, \ldots, s_{kw}\}$, 
e.g. $p(x_2, 1)=j$. 
The contraction functions $F_{ij}^{(k)}$ are introduced as  
\begin{eqnarray}
F_{ij}^{(k)}(1, j, 2^{-w})=(2, x_2, 2^{-3w}{\tilde r}_{ij}^{(k)}),
\end{eqnarray}
where $p(x_2, 2)=k$ and $p(x_2, 3)=i$, 
and the contraction ratio of $F_{ij}^{(k)}$ is $r_{ij}^{(k)}=2^{-2w}{\tilde r}_{ij}^{(k)}$. 
Each index $k$ indicates  
an allowed configuration of $\{s_{w+1}, s_{w+2}, \ldots, s_{2w}\}$ 
and the index $i$ indicates  
an allowed configuration of $\{s_{2w+1}, s_{2w+2}, \ldots, s_{3w}\}$. 
Thus a finite set of contraction functions is obtained  
corresponding to all the allowed generations 
from the configuration $j$ to $x_2$ where $p(x_2, 3)=i$. 
After the contractions by $F_{ij}^{(k)}$ 
we find a set of intervals, 
where each interval $[x_2, x_2+2^{-3w}{\tilde r}_{ij}^{(k)}]$   
corresponds to an allowed line configuration $\{s_1, \ldots, s_{3w}\}$ on the lattice.

When the number of rows is generally increased as $h\mapsto h+1$, 
$2w$ bonds will be added. 
The contraction function works as 
\begin{eqnarray}
F_{ij}^{(k)}(h, x_h, \Delta)=(h+1, x_{h+1}, 2^{-2w}{\tilde r}_{ij}^{(k)}\Delta),
\end{eqnarray}
where $x_{h+1}=x_h+2^{-(2h-1)w}k+2^{-2hw}i$, 
$p(x_h, 2h-1)=p(x_{h+1}, 2h-1)=j$, $p(x_{h+1}, 2h)=k$ and $p(x_{h+1}, 2h+1)=i$. 
This provides contractions of intervals 
\begin{eqnarray}
[x_h, x_h+\Delta]\mapsto [x_{h+1}, x_{h+1}+2^{-2w}{\tilde r}_{ij}^{(k)}\Delta].
\end{eqnarray}
Contraction functions $F_{ij}^{(k)}$ 
corresponding to the generation $h\to h+h_0$ ($h_0\in{\bf N}$)
can be introduced by straightforward generalizations.  

Taking the limit $h\to\infty$, 
we obtain fractal sets $\{E_j\}$ generated by the graph-directed IFS 
where the set of contraction functions is given by $\{F_{ij}^{(k)}\}$.  

As for the fractal dimension of graph-directed sets 
generated by a graph-directed IFS $\{F_{\alpha\alpha'}^{(\gamma)}\}$, 
the following results are generally obtained[\cite{B}, \cite{MW}  
and see for example \cite{E}, \cite{F3}]. 
Let us introduce a $q$-dimensional matrix $A^{(s)}$ 
\begin{eqnarray}
(A^{(s)})_{\alpha\alpha'}=\sum_{\gamma} (r_{\alpha\alpha'}^{(\gamma)})^s, 
\hspace{0.6cm}s\in{\bf R}, 
\end{eqnarray}
where $r_{\alpha\alpha'}^{(\gamma)}$ 
is the contraction ratio of $F_{\alpha\alpha'}^{(\gamma)}$ 
and $\alpha, \alpha'=1, 2, \ldots, q$. 
Let $\rho_\alpha(s)$ be the eigenvalues of $A^{(s)}$. 
From the Perron-Frobenius theorem, 
we know that 
there exists a real and positive eigenvalue $\rho_{\rm max}{(s)}$ 
which satisfies $|\rho_\alpha(s)|\leq\rho_{\rm max}(s)$ for all $\alpha$. 
Then, 
\begin{theorem}
Let $E_1$, $E_2$,\ldots, $E_q$ be a family of graph-directed sets 
generated by a graph-directed IFS $\{F_{\alpha\alpha'}^{(\gamma)}\}$ 
that satisfies the transitivity and the separation conditions. 
Then, \\
1) ${\dim}_{\rm H}E_\alpha={\dim}_{\rm B}E_\alpha$ $\:(\alpha=1, 2, \ldots, q)$,\\
2) there exists an positive number $s$ such that 
\begin{eqnarray}
{\dim}_{\rm H}E_1={\dim}_{\rm H}E_2=\cdots={\dim}_{\rm H}E_q=s, 
\label{seq}
\end{eqnarray}
3) the number $s$ is the unique solution of the equation 
\begin{eqnarray}
\rho_{\rm max}{(s)}=1. 
\label{rmax}
\end{eqnarray}
\end{theorem}

In the case of the six-vertex model, let 
\begin{eqnarray}
{\tilde r}_{ij}^{(k)}=\exp(-\beta_0\epsilon^{(k)}_{ij}), 
\label{bratio}
\end{eqnarray}
where $\beta_0$ is a non-negative constant 
and $\epsilon^{(k)}_{ij}$ is the sum of the energy of added $w$ vertices. 
The contraction ratio is 
$r_{ij}^{(k)}=(1/2)^{2w}{\tilde r}_{ij}^{(k)}$, 
and thus the matrix $A^{(s)}$ becomes $A^{(s)}=(1/2)^{2ws}V'$, 
where $V'$ is the row to row transfer matrix 
with the restricted $q$ basis. 
Because of the first factor $(1/2)^{2ws}$, 
the separation condition is satisfied. 

If the configuration $j$ and $i$ are $n$-equivalent,
the configuration $i$ can be generated from $j$
within $2n$ times operations of the transfer matrix.[\cite{n-eq_6V}]
Thus the transitivity condition corresponds to the fact that 
all the boundary configurations on the first row 
are $n$-equivalent to each other with some finite $n$: 
when we consider the set of all the $n$-equivalent boundary conditions,
the corresponding graph-directed IFS is transitive.

When we introduce all the admissible line configurations 
with fixed $m$ lines on the first row, 
these configurations form the complete set of $n$-equivalent boundary conditions. 
The transfer matrix is block diagonalized according to $m$. 
The fractal dimension $s$ is obtained from the condition (\ref{rmax}), i.e. 
\begin{eqnarray}
(\frac{1}{2})^{2ws}\lambda_{\rm max}(s\beta_0)=1, 
\end{eqnarray}
where $\lambda_{\rm max}(\beta)$ is the maximum eigenvalue 
of the block element of the transfer matrix, 
with the temperature $\beta=s\beta_0$. 
Hence we obtain the relation 
\begin{eqnarray}
s=\frac{-\beta f(\beta)}{2\log 2}, \hspace{1.2cm}
-\beta f(\beta)=\frac{1}{w}\log\lambda_{\rm max}(\beta),
\label{rel}
\end{eqnarray}
where $f(\beta)$ is the free energy of the six-vertex model 
with $m$ lines. 
We already know[\cite{n-eq}, \cite{n-eq_6V}] that
$n$-equivalent (i.e. transitive) 
boundary conditions yield the identical $f(\beta)$ (i.e. identical fractal dimension) 
in the thermodynamic limit, 
that is consistent with (\ref{seq}). 
Eq. (\ref{rel}) is the functional relation for $s$. 
The dimension $s$ is not identical with the multifractal dimension. 
The correspondences of the concepts 
in the theory of lattice models and those in the fractal geometry 
are summerised in Table 1.

The resulted $s$ depends on 
the 'anisotropy' parameter $\Delta$ of the six-vertex model 
only through the ratio $r_{ij}$, 
and hence the correspondence is valid for all $\Delta$. 

However 
$\lambda_{\rm max}$ has different functional form depending on $\Delta$. 
The maximum eigenvalue $\lambda_{\rm max}$ of the transfer matrix 
for the six-vertex model is (see for example [\cite{RJB}]) 
\begin{eqnarray}
\lambda_{\rm max}(\beta)=a^w\prod_{j=1}^wL(z_j)+b^w\prod_{j=1}^wM(z_j), 
\end{eqnarray}
where 
$L(z)=(ab+(c^2-b^2)z)/(a^2-abz)$, 
$M(z)=(a^2-c^2-abz)/(ab-b^2z)$, 
$a=e^{-\beta\epsilon_1}$, 
$b=e^{-\beta\epsilon_2}$ and 
$c=e^{-\beta\epsilon_3}$. 
The complex numbers $z_j$ satisfy the equations
\begin{eqnarray}
z_j^w=(-1)^{m-1}\prod_{k=1 (k\neq j)}^m\frac{s_{kj}}{s_{jk}}
\hspace{0.6cm}(j=1,2,\ldots ,m),
\label{eqforzj}
\end{eqnarray}
where $s_{ij}=1-2\Delta z_j+z_iz_j$, $\Delta=(a^2+b^2-c^2)/2ab$
and $m$ is the number of the lines. 
The same equations appear   
when we try to obtain the lowest energy of the XXZ quantum spin chain. 
If $1<\Delta$, it is known that $z_j$'s are real and $\lambda_{\rm max}=a^w+b^w$. 
If $\Delta<1$, $z_j$'s lie on the unit circle  in the complex plane. 
When one consider the limit $w\to\infty$ with fixed $m/w$, 
Eq. (\ref{eqforzj})  is reduced to an Fredholm  integral equation of the second kind,  
and its solution is obtained for $m/w=1/2$, with some assumptions.  
An additional structural transition of $\lambda_{\rm max}$ exists at $\Delta=-1$, 
where  an appropriate parameter $\mu$, $\Delta=-\cos\mu$, 
changes from real to pure imaginary. 

These changes correspond to the "phase transitions" of the six-vertex model, 
and of the XXZ quantum spin chain. 
This fact also means that 
the functional relation for the fractal dimension of this fractal set 
shows structural changeat $\Delta=1$ and $-1$  in the limit $w\to\infty$. 

If one takes $\epsilon^{(k)}_{ij}=0$ for all $i$, $j$ and $k$,  
then the partition function is equal to the number of
all the admissible nests of lattice paths for the fixed $m$. 
The fractal dimension $s$ becomes proportional to the entropy of the nests.  
This case, $\Delta=1/2$, 
corresponds to the high temperature limit of the six-vertex model. 

The case $\Delta=0$ 
is particularly simple and $\lambda_{\rm max}$ is expressed as 
\begin{equation}
\log\lambda_{\rm max}=-\beta\epsilon_1
+\int_{-\infty}^{+\infty}
\frac{\sinh(\frac{\pi}{2}+\gamma)x}{4x\cosh^2\frac{\pi}{2}x}dx,
\nonumber
\end{equation}
where
$a/c=\sin\frac{1}{2}(\frac{\pi}{2}-\gamma)$ and 
$b/c=\sin\frac{1}{2}(\frac{\pi}{2}+\gamma) $.
This case is the fractal set which corresponds to the XY spin chain. 
This case also concerns the tiling problem (the dimer problem) 
solved by Kasteleyn[\cite{Kast}] and Temperley and Fisher[\cite{TF}]. 
The solutions of tiling problems 
are classified by Cohn, Kenyon and Propp[\cite{CKP}] 
and also classified[\cite{n-eq_6V}] 
with the use of the $n$-equivalence.


 



\newpage 




\vspace{1.2cm}
\tabcolsep=0.8cm
\begin{tabular}{c|c} 
lattice models & fractal geometry \\ \hline
transfer matrix $V$ & matrix $A^{(s)}$ \\
$n$-equivalence & transitivity \\
free energy $f(\beta)$ & equation for the fractal dimension $s$
\end{tabular}

\vspace{0.6cm}
Table.1: \hspace{0.2cm}
Correspondence between lattice models and fractal geometry.

\vspace{2.4cm}
Figurecaptions:\\
\vspace{0.2cm}

Fig.1(a): \hspace{0.2cm}
Six vertices, corresponding lines and associated energies.\\

Fig.1(b):  \hspace{0.2cm}
An allowed line configuration and the transfer matrix $V$.\\

Fig.2:  \hspace{0.7cm}
Dots and directed edges which denote contraction functions.\\

Fig.3:  \hspace{0.7cm}
The rectangular lattice with $w$ lows and the order of bonds.

\newpage 
\begin{center}
\includegraphics[width=12cm,clip]{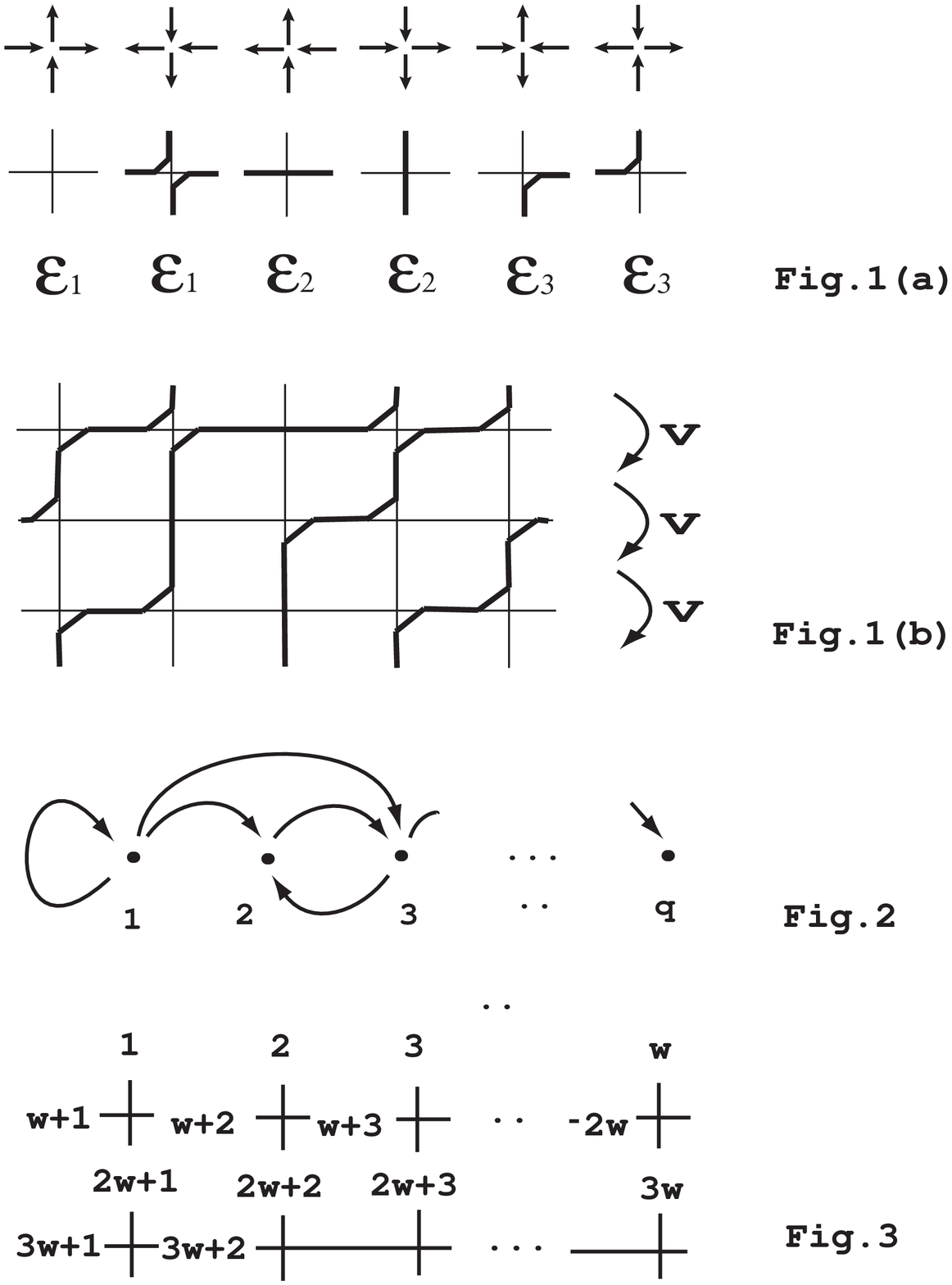} \\
\end{center}


\begin{thebibliography}{99}

\bibitem{L1} E.  H. Lieb, 
Exact Solution of the F Model of an Antiferromagnetic, 
Phys. Rev. Lett. 18 (1967) 1046.
\bibitem{L2} E. H. Lieb, 
Exact Solution of the Two-dimensional Slater KDP Model of a Ferroelectric, 
Phys. Rev. Lett. 19 (1967) 108.
\bibitem{L3} E. H. Lieb, 
Residual Entropy of Square Ice, 
Phys. Rev. 162  (1967) 162.
\bibitem{S} B. Sutherland, 
Exact Solution of a Two-dimensional Model for Hydrogen-bonded Crystals, 
Phys. Rev. Lett. 19 (1967) 103.

\bibitem{J} M. Jimbo (ed.), 
Yang-Baxter equation in Integrable Systems, 
World Scientific, 1990.
\bibitem{JM} M. Jimbo and T. Miwa, 
Algebraic Analysis of Solvable Lattice Models, 
CBMS Regional Conference Series in Mathematics 85, Amer. Math. Soc., 1995. 

\bibitem{HJKPS} T. C. Halsey, M. H. Jensen, L. P. Kadanoff, I. Procaccia and B. I. Shraiman, 
Fractal Measures and their Singularities: the Characterization of Strange Sets, 
Phys. Rev. A33  (1986) 1141.
\bibitem{EM} G. A. Edgar and R. D. Mauldin, 
Multifractal Decompositions of Digraph Recursive Fractals, 
Proc. London Math. Soc. (3) 65 (1992) 604. 

\bibitem{AKW} Y. Akutsu, A. Kuniba, M. Wadati, 
Exactly Solvable IRF Models. III. A New Hierarchy of Solvable Models, 
J. Phys. Soc. Jpn. 55 (1986) 1880. 
\bibitem{KAW} A. Kuniba, Y. Akutsu, M. Wadati, 
Exactly Solvable IRF Models. V. A Further New Hierarchy, 
J. Phys. Soc. Jpn. 55 (1986) 2605. 

\bibitem{n-eq} K. Minami, 
An Equivalence Relation of Boundary/Initial Conditions and the Infinite Limit Properties,  
J. Phys. Soc. Jpn. 74 (2005) 1640. 
\bibitem{n-eq_6V} K. Minami, 
The Free Energies of Six-Vertex Models and the n-Equivalence Relation, 
J. Math. Phys. 49 (2008)  033514.

\bibitem{B} T. Bedford, 
Dimension and Dynamics for Fractal Recurrent Sets, 
J. London Math. Soc. (2) 33 (1986) 89. 
\bibitem{MW} R. D. Mauldin and S. C. Williams, 
Hausdorff Dimension in Graph Directed Constructions, 
Trans. Amer. Math. Soc. 309 (1988) 811. 
\bibitem{E} G. A. Edgar, 
Measure, Topology, and Fractal Geometry, 
Springer-Verlag, 1990.
\bibitem{F3} K. J. Falconer, 
Techniques in Fractal Geometry, 
John Wiley and Sons, 1997. 

\bibitem{Kast} P. W. Kasteleyn, 
The Statistics of Dimers on a Lattice, 
Physica 27 (1961) 1209.
\bibitem{TF} H. N. V.Temperley and M. E. Fisher, 
The Dimer Problem in Statistical Mechanics - An Exact Result, 
Philos. Mag. 6 (1961) 1061.
\bibitem{CKP} H. Cohn, R. Kenyon and J. Propp, 
A Variational Principle for Domino Tilings, 
J. Amer. Math. Soc. 14 (2001) 297.

\bibitem{RJB} R.J.Baxter, 
Exactly Solved Models in Statistical Mechanics, 
Academic Press, 1982. 
\end{thebibliography}
\end{document}